# Stretching self-interacting, partially directed, flexible and semi-flexible polymers by an external force

Pui-Man Lam and Yi Zhen
Physics Department, Southern University
Baton Rouge, Louisians 70813

We study the model of a partially directed flexible or semi-flexible homopolymer on a square lattice, subject to an externally applied force, in a direction either parallel to, or perpendicular to the preferred direction. The polymer is self-interacting and can therefore undergo a collapse transition. We show that this model can be solved and we obtain the force-temperature phase diagrams which, for the case of flexible polymers, agree with that of Brak et al obtained using a different method. At sufficiently low temperatures, the polymer conformation changes from compact to coil state as the force is increased beyond a critical value. This transition is second or first order for the completely flexible or semi-flexible polymer, respectively.

## I. Introduction

The development of atomic force microscopy and optical and magnetic tweezers have allowed detailed studies of micro-manipulations of single polymers molecules [1-8]. This has led to renewed interest on the conformation problem of a polymer under an external applied stretching force [9-25]. A well-studied, exactly solvable model of polymer in a poor solvent is the self-interacting, partially directed self-avoiding walk model (IPDSAW). We will consider this model on the square lattice with the polymer oriented along the preferred direction, the horizontal direction and the other, perpendicular direction will be the vertical direction. In most cases studied, the externally applied force is along the horizontal direction. However, recently Kumar and Giri [15] considered pulling in both directions, with focus on finite size effects calculated numerically rather than thermodynamic transitions and exact results. Very recently, Rajesh et al. [24] and Brak et al. [25] obtained analytic results for the model with pulling in both directions. They obtained the critical force-temperature phase diagrams, but only for the case of a flexible polymer. Zhou et al [20] predicted that the addition of stiffness changes the order of the collapse transition to first order. This was verified by Owczarek and Prellberg [21] by solving exactly the semi-flexible case of IPDW.

For external force in the horizontal direction, Zhou et al [20] had applied the Lifson method and obtained results for both flexible and semi-flexible polymers. We had recently applied the same method to study adsorption of flexible and semi-flexible polymers pulled by a force in the horizontal direction, near an attractive wall [27]. By applying this method to the stretching problem in both directions, we obtained results for critical force-temperature phase diagram, for both the flexible and semi-flexible polymers. At sufficiently low temperatures, the polymer conformation changes from compact to coil state as the force is increased beyond a critical value. This transition is second or first order for the completely flexible or semi-flexible polymer, respectively. In section III we describe the method used in solving it and the results obtained. Section IV is the summary. Details of the calculations are given in the Appendix.

## II. The Model

The 2D partially directed walk of N steps on a square lattice is shown in Fig. 1, each step connecting two monomers. The model is partially directed in the sense that monomer units can be added only in the positive x-direction, while in the transverse direction, they can be added both in the positive and negative y-directions. The length of the bond connecting two consecutive monomers i and i+1 is fixed to $a_0$. If any two monomer i and i+m ($m \geq 3$) occupy nearest neighboring lattice sites, an attractive energy of magnitude ***e*** is gained. Usually real polymers are semi-flexible. We associate an energy penalty of magnitude $\Delta$ to each local direction change of the chain [15, 28]. In addition, the other end of the polymer can be pulled by a stretching forces $f_x$, $f_y$, applied on the last monomer, in the positive x and y directions respectively.

## III. The Method and Results

The partially directed model described above can be solved using the method of [20]. To calculate the free energy density of the polymer, a given configuration of the 2D chain is divided into a linear sequence of ***b***-sheet segments and coil segments using the Lifson approach [26, 28]. A ***b***-sheet segment is defined as a folded segment of $n_b \geq 2$ consecutive columns, in which contacting interactions exist between any two adjacent ones. Two consecutive ***b***-sheet segments are separated by a coil segment, which is a segment of $n_c \geq 0$ consecutive columns in which all monomers are free of contacts. For example, the configuration shown in Fig. 1 consists of one ***b***-sheet of 22 monomers and one coil of five monomers. After making such a distinction between ***b***-sheets and coils, we proceed by first calculating the partition functions of ***b***-sheets and coils separately.

We begin by considering the ***b***-sheet configurations. In Fig. 2a we show such a configuration with six columns connected by five horizontal steps. The numbers $l_i$ give the number of steps in the i-th column. The arrows give the direction of the step. $f_x$, $f_y$ are the forces that can be applied on the last monomer. Fig. 2b shows a ***b***-sheet configuration with exactly the same number of steps in each column as in Fig. 1a, except that now the directions of the arrows are reversed. It is clear that for an arbitrary ***b***-sheet configuration, a similar configuration with the same number of step in each column, but with the directions of the arrows reversed, is also allowed. Also in any $b$-sheet configuration, the directions of the steps in consecutive columns are reversed.

Defining $a = e^{e/T}$, $s = e^{-\Delta/T}$, $p = e^{f_x a_0 /T}$, $q = e^{f_y a_0 /T}$, where T is the temperature, and we have taken the Boltzmann constant to be unity, we can express the partition of a $b$-sheet consisting of n steps and $n_b$ columns as

$$Z_b(n) = \sum_{n_b=2}^{[(n+1)/2]} \sum_{\{l_i \geq 1\}} \prod_{i=1}^{n_b-1} a^{\min(l_i, l_{i+1})} s^{2(n_b-1)} p^{n_b} d_{l_1+l_2+\ldots l_{n_b}}^{n-n_b} \left[ q^{\sum_{i=1}^{n_b}(-1)^{i+1}l_i} + q^{\sum_{i=1}^{n_b}(-1)^{i}l_i} \right] \quad (1)$$

Here and in the following, the square brackets that appear in the upper summation limit, enclosing the ratio of two integers, denote the integer value of the ratio. $d_{l_1+l_2+\ldots l_{n_b}}^{n-n_b}$ is the Kronecker delta function, which is equal to unity if $n - n_b = l_1 + l_2 + \bullet\bullet\bullet + l_{n_b}$ and zero otherwise. The first three terms in the product are from the self-interaction, stiffness and force in horizontal direction, respectively. The two terms in the square brackets come from pulling in the vertical direction, each coming from configurations having the same number of steps in all columns, but with the directions of steps in reverse, as illustrated in Fig. 2a and Fig. 2b. The exponents $(-1)^{i+1}l_i$ and $(-1)^{i}l_i$ arise because the step directions in consecutive columns are reversed.

It is easier to calculate the generating function $G_b(z)$ of the partition function $Z_b(n)$ than to calculate $Z_b(n)$ directly,

$$G_b(z) \equiv \sum_{n=3}^{N} (z/a)^n Z_b(n)$$

$$= \sum_{n=3}^{N} \sum_{m=2}^{[(n+1)/2]} \sum_{\{l_j \geq 1\}} d_{l_1+l_2+\bullet\bullet\bullet+l_m}^{n-m} s^{2(m-1)} p^m \left(\frac{z}{a}\right)^{l_1+l_2+\bullet\bullet\bullet l_m} \left(\frac{z}{a}\right)^m$$

$$\otimes \prod_{i=1}^{m-1} a^{\min(l_i, l_{i+1})} \left[ \left(\frac{1}{q}\right)^{l_1} q^{l_2} \left(\frac{1}{q}\right)^{l_3} \bullet\bullet\bullet q^{(-1)^m l_m} + q^{l_1}\left(\frac{1}{q}\right)^{l_2} q^{l_3} \bullet\bullet\bullet q^{(-1)^{m-1} l_m} \right] \quad (2)$$

where $\zeta$ is an arbitrary parameter and N is the maximum number of steps. Using $\min(l_i, l_{i+1}) = [l_i + l_{i+1} - |l_i - l_{i+1}|]/2$, this can be written as

$$G_b(z) = \sum_{m=2}^{[(N+1)/2]} \sum_{\{l_j \geq 1\}} s^{2(m-1)} \left(\frac{pz}{a}\right)^m \prod_{i=1}^{m-1} \frac{z^{(l_i+l_{i+1})/2}}{a^{|l_i-l_{i+1}|/2}}$$

$$\otimes \left[ \left(\frac{1}{q}\right)^{l_1} q^{l_2} \left(\frac{1}{q}\right)^{l_3} \bullet\bullet\bullet q^{(-1)^m l_m} + q^{l_1}\left(\frac{1}{q}\right)^{l_2} q^{l_3} \bullet\bullet\bullet q^{(-1)^{m-1} l_m} \right]$$

$$= \sum_{m=2}^{[(N+1)/2]} \sum_{\{l_j \geq 1\}} s^{2(m-1)} \left(\frac{pz}{a}\right)^m \left(\frac{z}{a}\right)^{l_1/2} \left(\frac{z}{a}\right)^{l_m/2}$$

$$\otimes \left\{ q^{(-1)^m l_m} \left[ \left( \frac{q^{-l_1} \mathbf{z}^{(l_1+l_2)/2}}{a^{|l_1-l_2|/2}} \frac{q^{l_2} \mathbf{z}^{(l_2+l_3)/2}}{a^{|l_2-l_3|/2}} \right) \left( \frac{q^{-l_3} \mathbf{z}^{(l_3+l_4)/2}}{a^{|l_3-l_4|/2}} \frac{q^{l_4} \mathbf{z}^{(l_4+l_5)/2}}{a^{|l_4-l_5|/2}} \right) \bullet \bullet \bullet \right] \right.$$

$$\left. + q^{(-1)^{m-1} l_m} \left[ \left( \frac{q^{l_1} \mathbf{z}^{(l_1+l_2)/2}}{a^{|l_1-l_2|/2}} \frac{q^{-l_2} \mathbf{z}^{(l_2+l_3)/2}}{a^{|l_2-l_3|/2}} \right) \left( \frac{q^{l_3} \mathbf{z}^{(l_3+l_4)/2}}{a^{|l_3-l_4|/2}} \frac{q^{-l_4} \mathbf{z}^{(l_4+l_5)/2}}{a^{|l_4-l_5|/2}} \right) \bullet \bullet \bullet \right] \right\} \quad (3)$$

In each of the square brackets is a product of (m-1) terms, grouped in pairs by round brackets. For (m-1) an even integer, there are exactly (m-1)/2 pairs, while for (m-1) an odd integer, there are [(m-1)/2] pairs and one extra unpaired term.

We now define a matrix $\hat{\Lambda}(q)$ with the following elements:

$$\Lambda(q)_{ij} = \frac{q^i \mathbf{z}^{(i+j)/2}}{a^{|i-j|/2}} \quad (4)$$

The terms inside the round brackets in Eqn. (3) then become a matrix $\hat{\Gamma}(q) = \hat{\Lambda}(q)\hat{\Lambda}(q^{-1})$, or $\hat{\Gamma}(q^{-1}) = \hat{\Lambda}(q^{-1})\hat{\Lambda}(q)$, with elements

$$\Gamma(q)_{ij} = \sum_k \frac{q^{i-k} \mathbf{z}^{k+(i+j)/2}}{a^{|i-k|/2+|k-j|/2}}. \quad (5)$$

Furthermore, we define column vectors $\vec{x}$ and $\vec{u}(q)$ whose transpose are row vectors given by:

$$\vec{x}^T = \left( \left(\frac{\mathbf{z}}{a}\right)^{1/2}, \left(\frac{\mathbf{z}}{a}\right), \left(\frac{\mathbf{z}}{a}\right)^{3/2}, \bullet\bullet\bullet \right); \quad \vec{u}(q)^T = \left( q\left(\frac{\mathbf{z}}{a}\right)^{1/2}, q^2\left(\frac{\mathbf{z}}{a}\right), q^3\left(\frac{\mathbf{z}}{a}\right)^{3/2}, \bullet\bullet\bullet \right). \quad (6)$$

With these definitions, we can write the generating function $G_b(\mathbf{z})$ as:

$$G_b(\mathbf{z}) = \sum_{m=3,5,7,\bullet\bullet\bullet,odd}^{[(N+1)/2]} s^{2(m-1)} \left(\frac{ps}{a}\right)^m \vec{x}^T \left\{ \hat{\Gamma}(q)^{(m-1)/2} \vec{u}(q) + \hat{\Gamma}(q^{-1})^{(m-1)/2} \vec{u}(q^{-1}) \right\}$$

$$+ \sum_{m=2,4,6\bullet\bullet\bullet,even}^{[(N+1)/2]} s^{2(m-1)} \left(\frac{ps}{a}\right)^m \vec{x}^T \left\{ \hat{\Gamma}(q)^{[(m-1)/2]} \hat{\Lambda}(q) \vec{u}(q^{-1}) + \hat{\Gamma}(q^{-1})^{[(m-1)/2]} \hat{\Lambda}(q^{-1}) \vec{u}(q) \right\} \quad (7)$$

The terms in the second sum in Eqn. (7) contain an extra factor $\hat{\Lambda}(q)$ or $\hat{\Lambda}(q^{-1})$ because for m even, (m-1) is odd, resulting in an extra unpaired factor $\hat{\Lambda}$.

The sums in Eqn. (7) can now be carried out, resulting in the following expression:

$$G_b(\mathbf{z}) = s^2 \left(\frac{ps}{a}\right)^2 \vec{x}^T \left\{ \hat{\Lambda}(q)\vec{u}(q^{-1}) + \hat{\Lambda}(q^{-1})\vec{u}(q) \right\}$$

$$+ \frac{ps}{a} \vec{x}^T \left\{ \left[ \hat{I} - \left(\frac{s^6 p^2}{a^2} \hat{\Gamma}(q)\right)^K \right] \left[ \hat{I} - \frac{s^6 p^2}{a^2} \hat{\Gamma}(q) \right]^{-1} - \hat{I} \right\} \vec{u}(q)$$

$$+ \frac{ps}{a} \vec{x}^T \left\{ \left[ \hat{I} - \left(\frac{s^6 p^2}{a^2} \hat{\Gamma}(q^{-1})\right)^K \right] \left[ \hat{I} - \frac{s^6 p^2}{a^2} \hat{\Gamma}(q^{-1}) \right]^{-1} - \hat{I} \right\} \vec{u}(q^{-1})$$

$$+ \frac{p^2 s^4}{a^2} \vec{x}^T \left\{ \left[ \hat{I} - \left( \frac{s^6 p^2}{a^2} \hat{\Gamma}(q) \right)^K \right] \left[ \hat{I} - \frac{s^6 p^2}{a^2} \hat{\Gamma}(q) \right]^{-1} - \hat{I} \right\} \hat{\Lambda}(q) \vec{u}(q^{-1})$$

$$+ \frac{p^2 s^4}{a^2} \vec{x}^T \left\{ \left[ \hat{I} - \left( \frac{s^6 p^2}{a^2} \hat{\Gamma}(q^{-1}) \right)^K \right] \left[ \hat{I} - \frac{s^6 p^2}{a^2} \hat{\Gamma}(q^{-1}) \right]^{-1} - \hat{I} \right\} \hat{\Lambda}(q^{-1}) \vec{u}(q) \quad (8)$$

where $K = \left[ \frac{N-1}{4} \right] + 1$, in which the square brackets denote integer value of the ratio of two integers.

The derivation of the generating function for the coil segment partition function is given in the Appendix. It has the form:

$$G_c(\mathbf{z}) = A(p,q,s,y)/B(p,q,s,y), \qquad y = \frac{z}{a} \quad (9)$$

$$A(p,q,s,y) = \frac{1}{2} s^2 \{ 2q + 2s^2[qp - (q^2+1)]y + [2q - (q^2+1)p(1+2s^2)]y^2$$
$$+ p[4 - 2q(1-s^2) - ps^2(q^2+1)]y^3 + 2qp^2 s^2 y^4 \} \quad (10)$$

$$B(p,q,s,y) = q - (q^2 + 1 + pq)y + [(q^2+1)p(1-s^2) + q]y^2$$
$$+ pq[-(1-2s^2)y^3 + ps^4 y^4 + p^2 s^4 y^5] \quad (11)$$

It is easy to check that for q=1, the functions A and B become

$$A(p,1,s,y) = s^2(1-y)(1-y-ps^2 y^2)(1+py) \quad (12)$$
$$B(p,1,sy) = (1-y-ps^2 y^2)[1-(1+p)y + p(1-s^2)y^2 - p^2 s^2 y^3] \quad (13)$$

The coil generating function then reduces to

$$G_c(\mathbf{z}) \xrightarrow[q=1]{} \frac{s^2(1-y)(1+py)}{1-(1+p)y + p(1-s^2)y^2 - p^2 s^2 y^3} \quad (14)$$

which agrees with the result of [20] for the case with no vertical force.

The generating function of whole polymer can be obtained by the summation

$$G(\mathbf{z}) = G_b(\mathbf{z}) + G_b G_c + (G_b G_c)^2 + (G_b G_c)^3 + \bullet\bullet\bullet + G_b[G_c G_b + (G_c G_b)^2 + \bullet\bullet\bullet]$$

$$+ G_c(\mathbf{z}) + G_c G_b + (G_c G_b)^2 + (G_c G_b)^3 + \bullet\bullet\bullet + G_c[G_b G_c + (G_b G_c)^2 + \bullet\bullet\bullet]$$

$$= G_b + \frac{G_b G_c + G_b G_c G_b}{1 - G_b G_c} + G_c + \frac{G_c G_b + G_c G_b G_c}{1 - G_b G_c}$$

$$= \frac{G_b(1+G_c) + G_c(1+G_b)}{1 - G_b G_c} \qquad (15)$$

The partition function Z(N) is related to the free energy density $g(f_x, f_y, T)$ of the polymer by $Z(N) = \exp[-Nbg(f_x, f_y, T)]$. Therefore in the thermodynamic limit $N \to \infty$, the free energy density is

$$g(f_x, f_y, T) = -e + T \ln z_0 \qquad (16)$$

where $z_0$ is the smallest positive root in the range $0 < z_0 < 1$ of the following equation:

$$1 - G_b(z_0) G_c(z_0) = 0 \qquad (17)$$

If Eqn.(17) has no root in the range $0 < z_0 < 1$, then $z_0 = 1$. The polymer's relative extensions $R_x$ and $R_y$ in the horizontal and vertical directions are

$$R_x = -\frac{\partial \ln z_0}{\partial \ln p}; \quad R_y = -\frac{\partial \ln z_0}{\partial \ln q} \qquad (18)$$

When both temperature T and external forces $f_x, f_y$ are sufficiently low, Eqn.(17) has no root in the range $0 < z_0 < 1$. Then the free energy density $g(f_x, f_y, T) = -e$, since $z_0 = 1$. There is no entropy contribution. The relative extensions $R_x$ and $R_y$ are zero by Eqns. (18). The polymer is in the collapsed $b$-sheet state. As the temperature or the external force is elevated to a certain point such that Eqn. (17) is satisfied exactly at $z_0 = 1$, a phase transition occurs. At this point the $b$-sheet phase changes to the extended phase.

We have numerically solved Eqn. (17) at $z_0 = 1$, using Eqn. (8) and (9) at fixed temperatures and obtained the critical force-temperature phase diagrams. We have done that for the special cases $f_x = 0$ and $f_y = 0$ separately. To obtain the $b$-sheet generating function using Eqn. (8), a finite value N for the number of steps must be used. We find that the results do not change when N is larger than 100. In our calculation we have actually checked that the solutions to Eqn. (17) do not change significantly for N values up to 500.

In Figs. 3a and 3b we show the critical force versus temperature diagrams for $f_y = 0$, for flexible ($\Delta = 0$) and semi-flexible ($\Delta = 0.25e$) polymers, respectively. The regions below and above the curve correspond to the collapsed and extended states respectively. By comparing Figs. 3a and 3b, it is seen that the stiffness in the chain tend to increase both the critical force and critical temperature.

In Figs. 4a and 4b we show the critical force-temperature diagrams for $f_x = 0$, for flexible ($\Delta = 0$) and semi-flexible ($\Delta = 0.25e$) polymers, respectively. The external force is now in the vertical direction. In the case of the flexible polymer we also show the result of Brak et al [25] which agrees with our result. Note that in both flexible and semi-flexible polymers, as the temperature approaches the critical value for no pulling force, the slope of the curve diverges in contrast to the analogous horizontal pulling curve. The chain stiffness has also the effect of increasing the critical force and critical temperature.

## IV Summary

We have studied the polymer model of partially directed walk with self-interaction, pulled by external forces both in the preferred horizontal and transverse vertical directions. The polymer can be either flexible ($\Delta = 0$), or semi-flexible ($\Delta > 0$). We show that this model can be analyzed and numerically solved. We obtained the critical force-temperature phase diagrams for the special cases $f_x = 0$ and $f_y = 0$ respectively. For the flexible polymer, our results agree with those of Brak et al [25], obtained using a different method. The chain stiffness has the effect of increasing the critical force and critical temperature.

Actually chain stiffness has a dramatic effect on the critical behavior of a polymer pulled by an external force. Zhou et al. [20] had shown by analytical arguments that for pulling in the preferred direction, any amount of stiffness in the chain would change the collapse transition from second order to first order. Although we have not studied this explicitly here, we do not expect the finding of ref. [20] on the effect of chain stiffness on the order of the collapsed transition to be modified for the case of pulling in the transverse direction.

## Appendix

In this appendix we derive the expression for the coil generating function. The configurational energy of a coil segment

$$E_c = m_c \Delta - n_c f_x a_0 + n_y f_y a_0 \qquad (A1)$$

where $m_c$ is the total number of bends, $n_c$ is the number of column, and $n_y$ is the number of steps in positive y direction. To calculate the partition function $Z_c(n)$ of a coil segment of n monomers, one needs to distinguish among four different boundary conditions.

(i) Both the left-most and the right-most column contain only one monomer. The partition function for such a situation is denoted as $Z_c^{1,1}(n,n_c)$.

(ii) The left-most column contains only one monomer, while the right-most column contains two or more monomers. The steps in the right-most column can be in either the +y or –y direction. The partition functions for the two directions are denoted as $Z_{c+}^{1,2}(n,n_c)$ and $Z_{c-}^{1,2}(n,n_c)$ respectively.

(iii) The left-most column contains two or more monomers, while the right-most column contains just one monomer. The partition function for such a situation is denoted as $Z_c^{2,1}(n,n_c)$.

(iv) Both the left-most and the right-most column contain two or more monomers. The steps in the right-most column can be in either the +y or –y direction. The partition functions for the two directions are denoted as $Z_{c+}^{2,2}(n,n_c)$ and $Z_{c-}^{2,2}(n,n_c)$ respectively.

We can write down the following iteration equations for the above defined six partition functions:

$$Z_c^{1,1}(n,n_c) = pZ_c^{1,1}(n-1,n_c-1) + ps[Z_{c+}^{1,2}(n-1,n_c-1)$$
$$+ Z_{c-}^{1,2}(n-1,n_c-1)], \quad (n \geq 2, 2 \leq n_c \leq n) \qquad (A2)$$

$$Z_{c+}^{1,2}(n,n_c) = qZ_{c+}^{1,2}(n-1,n_c) + psqZ_c^{1,1}(n-2,n_c-1) +$$
$$ps^2qZ_{c+}^{1,2}(n-2,n_c-1), (n \geq 3, 2 \leq n_c \leq n-1) \qquad (A3)$$

$$Z_{c-}^{1,2}(n,n_c) = rZ_{c-}^{1,2}(n-1,n_c) + psrZ_c^{1,1}(n-2,n_c-1) +$$
$$ps^2rZ_{c-}^{1,2}(n-2,n_c-1), (n \geq 3, 2 \leq n_c \leq n-1) \qquad (A4)$$

$$Z_c^{2,1}(n,n_c) = pZ_c^{2,1}(n-1,n_c-1) + ps[Z_{c+}^{2,2}(n-1,n_c-1)$$
$$+ Z_{c-}^{2,2}(n-1,n_c-1)], (n \geq 2, 2 \leq n_c \leq n-1) \qquad (A5)$$

$$Z_{c+}^{2,2}(n,n_c) = qZ_{c+}^{2,2}(n-1,n_c) + psqZ_c^{2,1}(n-2,n_c-1) +$$
$$ps^2qZ_{c+}^{2,2}(n-2,n_c-1), (n \geq 4, 2 \leq n_c \leq n-2) \qquad (A6)$$

$$Z_{c-}^{2,2}(n,n_c) = rZ_{c-}^{2,2}(n-1,n_c) + psrZ_c^{2,1}(n-2,n_c-1) +$$
$$ps^2rZ_{c-}^{2,2}(n-2,n_c-1), (n \geq 4, 2 \leq n_c \leq n-2) \qquad (A7)$$

where $r=1/q$.

The initial conditions are: $Z_c^{1,1}(1,n_c) = pq\mathbf{d}_{n_c}^1$; $Z_{c\pm}^{1,2}(1,n_c) = Z_{c\pm}^{1,2}(2,n_c) = 0$;

$Z_{c\pm}^{2,1}(1,n_c) = Z_{c\pm}^{2,1}(2,n_c) = 0$; and $Z_{c\pm}^{2,2}(1,n_c) = 0$, $Z_{c+}^{2,2}(2,n_c) = \frac{1}{2}pq\mathbf{d}_{n_c}^1$,

$Z_{c-}^{2,2}(2,n_c) = \frac{1}{2}pr\mathbf{d}_{n_c}^1$.

By iterating the above equations, the partition functions $Z_c^{1,1}(n,n_c)$, $Z_c^{2,1}(n,n_c)$, $Z_c^{1,2}(n,n_c) = Z_{c+}^{1,2}(n,n_c) + Z_{c-}^{1,2}(n,n_c)$ and $Z_c^{2,2}(n,n_c) = Z_{c+}^{2,2}(n,n_c) + Z_{c-}^{2,2}(n,n_c)$ can be calculated for finite n. The generating functions $G_c^{i,j}(\mathbf{z})$ of $Z_c^{i,j}(n,n_c)$, i,j =1or 2 defined as

$$G_c^{i,j}(\mathbf{z}) = \sum_{n=1}^{\infty} \left(\frac{\mathbf{z}}{a}\right)^n \sum_{n_c=1}^{n} Z_c^{i,j}(n,n_c) \qquad (A8)$$

can then be obtained:

$$G_c^{1,1}(\mathbf{z}) = \frac{py[q-(q^2+1)y+(-(q^2+1)ps^2+1)y^2+qps^2(2y^3+ps^2y^4)]}{B(p,q,s,y)} \qquad (A9)$$

$$G_c^{1,2}(\mathbf{z}) = \frac{p^2 sy^3[(q^2+1) - 2yq - 2ps^2 qy^2]}{B(p,q,s,y)} \tag{A10}$$

$$G_c^{2,1}(\mathbf{z}) = \frac{1}{2}\frac{y^3 p^2 s[(q^2+) - 2yq - 2ps^2 qy^2]}{B(p,q,s,y)} \tag{A11}$$

$$G_c^{2,2}(\mathbf{z}) = \frac{py^2}{2}\frac{q^2 + 1 - [2q + (q^2+1)p]y + 2p(1-s^2)qy^2 + 2p^2 s^2 qy^3}{B(p,q,s,y)} \tag{A12}$$

where $B(p,q,s,y)$ is given in Eqn. (11). One can easily check that the above equations reduce to those of ref. [20] in the case $q=1$, when there is no vertical force. By Taylor expanding the $G_c^{i,j}(\mathbf{z})$ in $\mathbf{z}$ one can check that the results agree with those obtained by solving Eqns. (A5)-(A7) to all orders in $\mathbf{z}$, so that they are indeed exact solutions.

The total partition function for a coil segment of n monomers is

$$Z_{bc}(n) = 2s^2 \sum_{n_c=1}^{n} Z_c^{1,1}(n,n_c) + s^3 \sum_{n_c=2}^{n-1} Z_c^{1,2}(n,n_c) + 2s^3 \sum_{n_c=2}^{n-1} Z_c^{2,1}(n,n_c) + s^4 \sum_{n_c=2}^{n-1} Z_c^{2,2}(n,n_c) \tag{A13}$$

while $Z_c(0) \equiv s^2$. The factor 2 in the first and third term of (A13) is due to the fact that, the first column of a $\mathbf{b}$-sheet following a coil segment of type (i) and (iii) can choose between two orientations. The generating function for the coil segment partition function is

$$\begin{aligned}
G_c(\mathbf{z}) &= \sum_{n=0}^{\infty} \left(\frac{\mathbf{z}}{a}\right)^n Z_c(n) \\
&= s^2 + 2s^2 G_{bc}^{1,1}(\mathbf{z}) + s^3 G_{bc}^{1,2}(\mathbf{z}) + 2s^3 G_{bc}^{2,1}(\mathbf{z}) + s^4 G_{bc}^{2,2}(\mathbf{z}) \\
&= \frac{A(p,q,s,y)}{B(pq,s,y)}
\end{aligned} \tag{A14}$$

as given by Eqn. (9).

Acknowledgement. Research supported by the Louisiana Board of Regents Support Fund Contract Number LEQSF(2007-10)-RD-A-29. We would like to thank Haijun Zhou for a critical reading of the original manuscript and very useful discussions. P.M. Lam would like to thank the hospitality of the Institute of Theoretical Physics, Beijing, China where part of this work was carried out.

REFERENCES

[1] For a brief review, see C. Bustamante, Z. Bryant and S.B. Smith, *Nature* **421**, 423 (2003)
[2] T. Strick, J.F. Allemand, V. Croquette and D. Bensimon, Physics Today **54**, 46 (2001)
[3] C. Danilowicz, C.H. Lee, V.W. Coljee and M. Prentiss, Phys. Rev. **E75**, 030902(R) (2007)
[4] J.E. Bemis, B.B. Akhremitchev and G.C. Walker, Langmuir **15**, 2799 (1999)
[5] B.J. Haupt, T.J. Senden and E.M. Sevick, Langmuir **18**, 2174 (2002)
[6] N. Gunari, A.C. Balazs and G.C. Walker J. Am. Chem. Soc. **129**, 10046 (2007)
[7] Y. Seol, G.M. Skinner and K. Visscher, A. Buhot and A. Halperin, Phys. Rev. Lett. **98**, 158103 (2007)
[8] C. Ke, M. Humeniuk, H. S-Gracz and P.E. Marszalek, Phys. Rev. Lett. **99**, 018302 (2007)
[9] H. Clausen-Schaumann, M. Seitz, R. Krautbauer and H.E. Gaub, Curr. Opin. Chem. Biol. **4**, 524 (2000)
[10] C. G. Baumann, V.A. Bloomfield, S. B. Smith, C. Bustamante, M.D. Wang and S.M. Block, Biophys. J. **78**, 1965 (2000)
[11] A. Halperin and E.B. Zhulina, Europhys. Lett. **15**, 417 (1991)
[12] P. Grassberger and H.P. Hsu, Phys. Rev. **E65**, 031807 (2002)
[13] D. Marenduzzo, A. Maritan, A. Rosa and F. Seno, Phys. Rev. Lett. **90**, 088301 (2003)
[14] A. Rosa, D. Marenduzzo, A. Maritan and F. Seno, Phys. Rev. **E67**, 041802 (2003)
[15] S. Kumar and D. Giri, Phys. Rev. **E72**, 052901 (2005)
[16] E.Orlandini, M. Tesi and S. Whittington, J. Phys. **A37**, 1535 (2004)
[17] S. Kumar, I. Jensen, J.L. Jacobsen and A. J. Guttmann, Phys. Rev. Lett. **98**, 128101 (2007)
[18] S. Kumar and G. Mishra, Phys. Rev. **E78**, 011907 (2008)
[19] J. Krawczyk, I. Jensen, A. L. Owczarek, and S. Kumar, Phys.Rev. E **79**, 031912 (2009).
[20] H. Zhou, Jie Zhou, Z.-C. Ou-Yang and S. Kumar, Phys. Rev. Lett. **97**, 158302 (2006)
[21] A.L. Owczrak and T. Prellberg, J. Stat. Mech. : Theor. Exp. P11010 (2007)
[22] I.R. Cooke and D.R.M. Williams, Europhys. Lett. **64**, 267 (2003)
[23] G. Mishra, D. Giri and S. Kumar, Phys.Rev. **E** 79, 031930(2009); Cond. Matt. 0907.1950v1
[24] R. Rajesh, G. Giri, I. Jensen and S. Kumar, Phys. Rev. **E78**, 021905 (2008)
[25] R. Brak, P. Dyke, J. Lee, A.L. Owczarek, T. Prellberg, A. Rechnitzer and S. G. Whittington, J. Phys. A42 ,085001 (2009); Cond. Matt. 0811.1392v1
[26] S. Lifson, J. Chem. Phys. **40**, 3705 (1964)
[27] P.M. Lam, Y. Zhen, H. Zhou and J. Zhou, Phys. Rev. **E79**, 061127 (2009)
[28] S. Doniach, T. Garel and H. Orland, J. Chem. Phys. **105**, 1601 (1996)

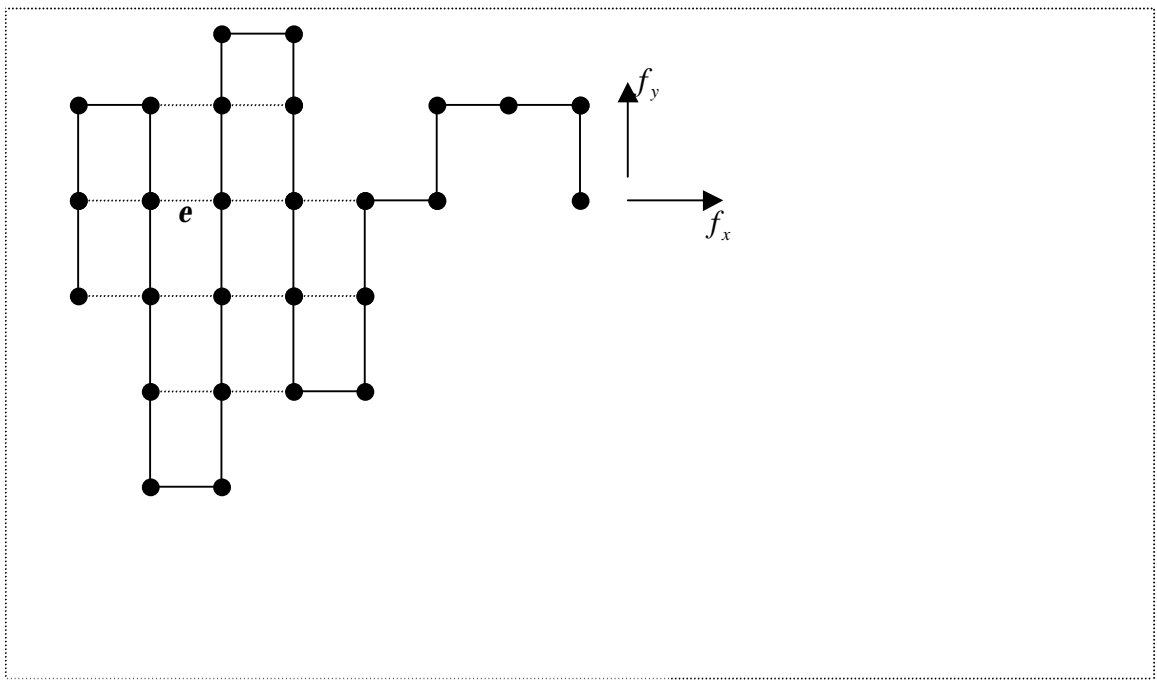

Fig. 1. Configuration of a polymer consisting of a $b$-sheet segment of 21 steps and a coil segment of 5 steps. Solid dots represent the monomers and dashed lines represent self-interaction $e$. The right end of the polymer is subjected to forces $f_x$ and $f_y$.

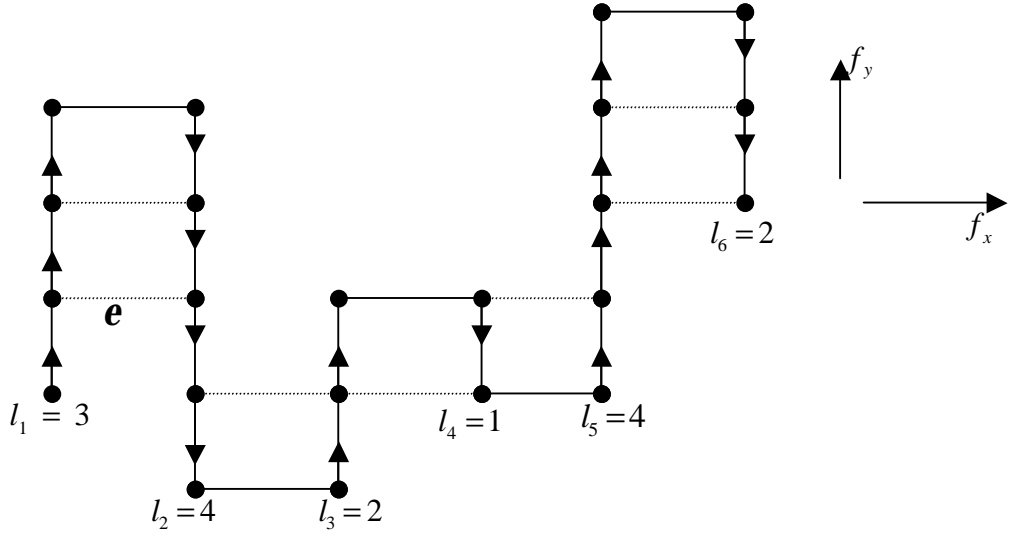

Fig. 2a. A **b**-sheet consisting of 21 steps. The numbers $l_i$ denotes the number of steps in the i-th column. The arrows give the direction of the steps. The directions of steps in consecutive columns are in reversed direction.

Fig. 2b. ***b***-sheet configuration with exactly the same number of steps $l_i$ as in Fig. 2a, but with the step directions reversed.

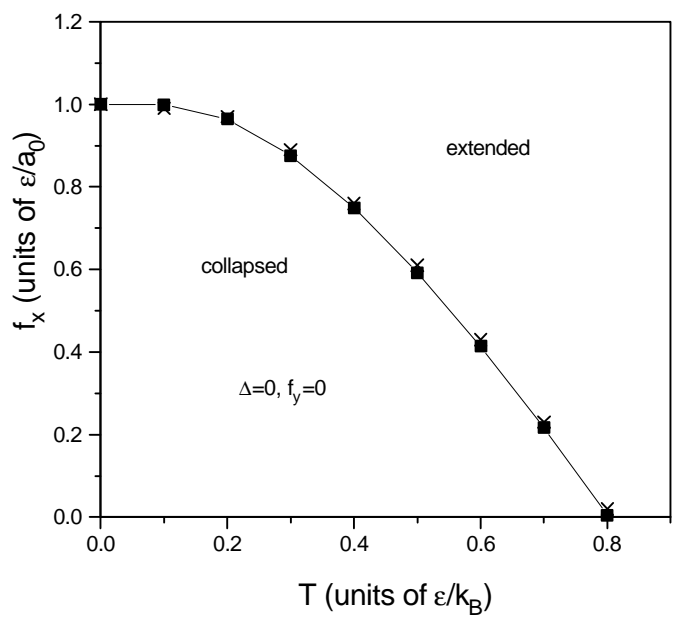

Fig. 3a. Phase diagram for flexible polymer ($\Delta=0$) at $f_y = 0$. The crosses denote result of ref [26].

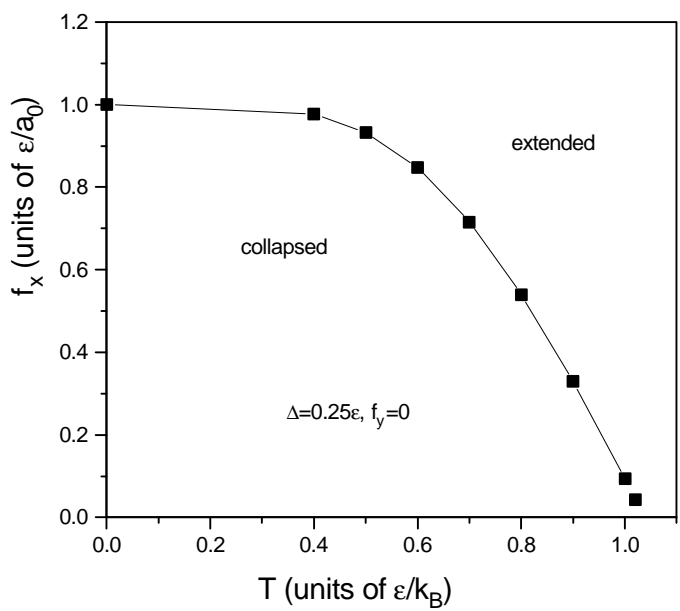

Fig. 3b. Phase diagram for semi-flexible polymer ($\Delta=0.25$) at $f_y=0$.

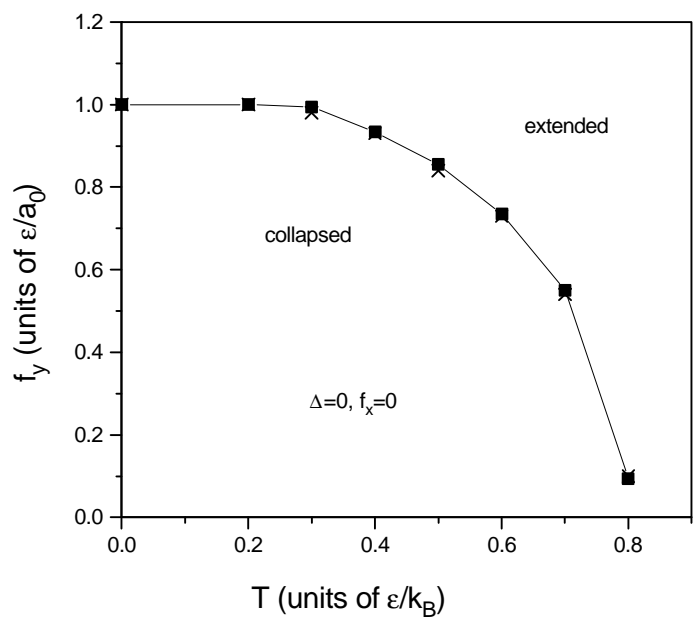

Fig. 4a. Phase diagram for flexible polymer ($\Delta=0$) at $f_x = 0$. Crosses are results of ref. [25].

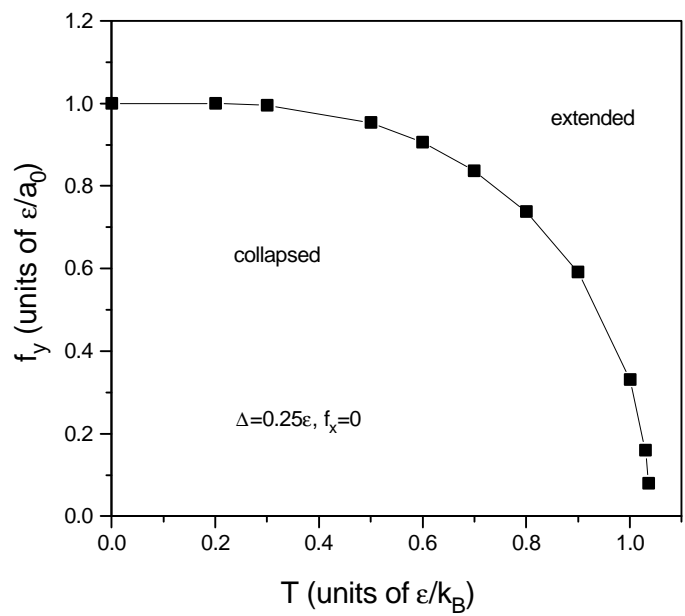

Fig. 4b. Phase diagram for semi-flexible polymer ($\Delta=0.25$) at $f_x = 0$.